\documentclass[pra,english,reprint,nofootinbib,aps,superscriptaddress,showpacs,showkeys]{revtex4-1}

\usepackage{babel,calc,amsmath,amsthm,amssymb,graphicx,subfigure,xcolor,mathtools,
      longtable,braket,bm,extarrows}

\usepackage[T1]{fontenc}
\usepackage[unicode=true]{hyperref}

\hypersetup{
     colorlinks=true,       		
     linkcolor=blue,          	
     citecolor=red,            
     urlcolor=magenta,           	
}
\newtheorem*{theorem*}{Theorem}
\newtheorem{theorem}{Theorem}

\newtheorem*{corollary*}{Corollary}

\newtheorem*{lemma*}{Lemma}

\newtheorem*{proposition*}{Proposition}

\newtheorem*{conjecture*}{Conjecture}
\theoremstyle{definition}

\newtheorem*{definition*}{Definition}
\theoremstyle{remark}

\newtheorem*{remark*}{Remark}

\newcommand{\Rom}[1]{\MakeUppercase{\romannumeral #1}}

\renewcommand{\figureautorefname}{Fig.}
\renewcommand{\figurename}{Fig.}

\begin{document}
\renewcommand{\figurename}{FIG.}
	\renewcommand{\figureautorefname}{FIG.}
   \title{Equality Conditions for an Additive Three-Observable Uncertainty Relation}

\author{Yao-Yi Zeng}
\affiliation{Theoretical Physics Division, Chern Institute of Mathematics and LPMC, Nankai University, Tianjin 300071, People's Republic of China}

\author{Zhi-Jie Liu}
\affiliation{Theoretical Physics Division, Chern Institute of Mathematics and LPMC, Nankai University, Tianjin 300071,
People's Republic of China}

\author{Jie Zhou}
\affiliation{College of Physics and Materials Science, Tianjin Normal University, Tianjin 300387, People's Republic of China}

\author{Jing-Ling~Chen}
\email{chenjl@nankai.edu.cn}
\affiliation{Theoretical Physics Division, Chern Institute of Mathematics and LPMC, Nankai University, Tianjin 300071, People's Republic of China}

   \date{\today}

\begin{abstract}
Uncertainty relations play a fundamental role in quantum mechanics by
quantifying the intrinsic limitations on the simultaneous sharpness of
incompatible observables. Beyond the standard two-observable product
form, additive uncertainty relations for triples of observables provide
a natural framework for describing collective constraints among three
noncommuting components. In this work, we study an additive
uncertainty relation for three Hermitian observables from the viewpoint
of rotational symmetry and covariance geometry. We give a short
rotational derivation by rotating the observable triple and applying the
Robertson uncertainty relation to the two transverse observables. This
derivation makes the saturation mechanism transparent and leads to a
necessary and sufficient condition for equality for general density
operators. In the nontrivial equality case, the covariance ellipsoid of
the observable triple degenerates into a disk perpendicular to the
expectation value of the commutator vector. We also discuss an inverse
construction based on finite-dimensional representations of the Lie algebra
\(\mathfrak{su}(2)\), which provides a systematic way to construct
observable triples with prescribed saturating states. These results
clarify the geometric and representation-theoretic structure underlying
the tightness of additive three-observable uncertainty relations.
\end{abstract}

\maketitle

\section{Introduction}

Uncertainty relations are among the most fundamental structural statements of
quantum mechanics. They quantify the intrinsic limitations on the simultaneous
sharpness of incompatible observables and distinguish quantum theory from
classical theories at the level of measurement statistics.
The uncertainty principle was first introduced by Heisenberg \cite{Heisenberg1927}, and
its standard deviation form for position and momentum was derived by Kennard
\cite{Kennard1927}.
It is also closely connected with the mathematical foundation of quantum mechanics, including Weyl's formulation of quantum theory \cite{Weyl1928}.
Since then, uncertainty relations have become a
central tool in quantum foundations and quantum information theory, with developments including entropic uncertainty relations
\cite{Deutsch1983,MaassenUffink1988}, uncertainty relations with quantum memory
\cite{Berta2010}, and various modern formulations \cite{Coles2017,Ozawa2003, BuschLahtiWerner2013, BuscemiHallOzawaWilde2014}. 

The best known variance-based form is the Heisenberg--Robertson uncertainty
relation \cite{Robertson1929}. For two Hermitian observables $A$ and $B$ and
a pure state $|\psi\rangle$, it takes the form
\begin{equation}
\Delta_\psi A\Delta_\psi B
\ge
\frac{1}{2}\left|\langle\psi|[A,B]|\psi\rangle\right|,
\label{eq:robertson_intro}
\end{equation}
where $[A,B]=AB-BA$, and $\Delta_\psi X$ denotes the standard deviation of observable $X$
in the state $|\psi\rangle$, namely
$(\Delta_\psi X)^2=\langle\psi|X^2|\psi\rangle-\langle\psi|X|\psi\rangle^2$. Schr\"odinger later generalized the Robertson relation by  introducing an additional covariance term involving the anti-commutator \cite{Schrodinger1930}. Although these product-form uncertainty relations provide fundamental bounds determined by the incompatibility of two observables, their lower bounds may become trivial for certain quantum states. This limitation motivates the study of additive uncertainty relations, where the collective incompatibility among observables can be characterized through sums of variances \cite{Huang2012,Chen2015,Song2017,MacconePati2014,Dodonov2018}.

For quantum systems involving more than two observables, additive uncertainty relations provide a natural generalization of the two-observable scenario.
In many physical systems, the relevant observables appear not merely
as a pair but as a triple or, more generally, as components of a larger noncommuting set.
A notable example is the Schrodinger triple $(p,q,r)$ studied by Kechrimparis and Weigert, where $q$
and $p$ form a canonical position--momentum pair and the third observable is
defined by $r=-p-q$; this triple satisfies the pairwise commutation relations
$[q,p]=[p,r]=[r,q]=\mathrm{i}\hbar$, with $\hbar$ denoting the reduced Planck
constant, and gives rise to uncertainty relations including the additive
relation \cite{PhysRevA.90.062118},
\begin{equation}
\Delta p^2+\Delta q^2+\Delta r^2\geq \sqrt{3}\hbar .
\label{eq:KW_intro}
\end{equation}
For spin-$1/2$ systems, Ma et al. investigated the three angular momentum components
and experimentally verified a corresponding additive uncertainty relation
\cite{PhysRevLett.118.180402}:
\begin{equation}
\begin{aligned}
(\Delta S_x)^2+(\Delta S_y)^2+(\Delta S_z)^2
\ge \\
\frac{1}{\sqrt{3}}\Bigl(
|\langle S_x\rangle|
+|\langle S_y\rangle|
+|\langle S_z\rangle|
\Bigr),
\end{aligned}
\label{eq:spin_additive_intro}
\end{equation}
where $S_x,S_y,S_z$ denote the three spin components,
$\Delta S_x,\Delta S_y,\Delta S_z$ their respective standard deviations,
and $\langle S_x\rangle,\langle S_y\rangle,\langle S_z\rangle$ their
corresponding expectation values in the given quantum state.

Similar additive uncertainty relations have been explored for other physical systems, including the three components of canonical momentum and observables associated with the hydrogen atom \cite{LIU2025170179,Liu_2025}.
These examples point to a common
structure: for a triple of Hermitian observables, the sum of their variances can
be bounded from below by expectation values of the corresponding commutator
operators.
A general form of such three-observable uncertainty relations
was established by Song and Qiao \cite{SONG20162925}.
However, while the inequality itself is known, its saturation structure has not been
fully characterized. This gap is relevant because equality conditions identify
the states and observable configurations for which an uncertainty bound is
tight.
A complete characterization of this saturation mechanism for general triples of observables is still lacking.

In this work, we investigate the tight additive uncertainty relation for three arbitrary Hermitian observables from the perspective of rotational symmetry and covariance geometry. We first provide a concise rotational derivation of the uncertainty bound by transforming the observable triple such that the expectation value of the commutator vector is aligned with a single coordinate axis, after which the Robertson relation for the transverse observables directly yields the additive bound. This approach not only proves the inequality but also reveals the underlying mechanism responsible for saturation.
Furthermore, we establish a necessary and sufficient condition for equality for general density operators. In the nontrivial equality case, we show that the covariance ellipsoid of the observable triple is reduced to a disk perpendicular to the expectation-value direction of the commutator vector, providing a geometric interpretation of tight uncertainty relations, such as those in Ref.\cite{LIU2025170179,Liu_2025}. We also present an inverse construction based on finite-dimensional representations of $\mathfrak{su}(2)$, which gives a systematic way to construct observable triples with prescribed saturating states.
Related applications of three-observable uncertainty relations to state identification have been considered in Ref.~\cite{Liu:2025qqm}.
Our results provide a unified characterization of tight three-observable uncertainty relations and clarify the connection between uncertainty bounds, covariance geometry, and quantum state structure.


The paper is organized as follows. Section~\Rom{2} presents the rotational proof
and derives the necessary and sufficient equality condition. Section~\Rom{3} discusses
the geometric interpretation of the equality condition using the covariance matrix and
the covariance ellipsoid. Section~\Rom{4} presents the inverse construction
from prescribed states based on finite-dimensional representations of
$\mathfrak{su}(2)$. Section~\Rom{5} contains the discussion and conclusion.
Appendix~A reviews the Robertson uncertainty relation and its equality
condition for general density operators. Appendix~B summarizes the covariance
matrix and covariance ellipsoid formalism.

\section{The additive uncertainty relation and its equality condition}
\label{sec:additive_relation}

 The main result of this section is the characterization of the equality condition for the additive uncertainty relation considered below. More precisely, a necessary and sufficient condition for its saturation is obtained for general density operators.

We consider three arbitrary Hermitian operators $L_1,L_2$  and $L_3$.
Since the commutator of two Hermitian operators is anti-Hermitian, one may always
introduce three Hermitian operators $V_1,V_2,V_3$ through
\begin{equation}
[L_1,L_2]=\mathrm{i}V_3,\qquad
[L_2,L_3]=\mathrm{i}V_1,\qquad
[L_3,L_1]=\mathrm{i}V_2.
\label{eq:commutator_form}
\end{equation}
For a quantum state $\rho$, we define
\begin{equation}
(\Delta_\rho A)^2=\langle A^2\rangle_\rho-\langle A\rangle_\rho^2,
\end{equation}
where $\langle A\rangle_\rho=\operatorname{tr}( \rho A)$ for an operator $A$.

For any quantum state $\rho$, one has
\begin{equation}
\begin{aligned}
(\Delta_\rho L_1)^2+(\Delta_\rho L_2)^2+(\Delta_\rho L_3)^2
\ge \\
\frac{1}{\sqrt{3}}\Bigl(
|\langle V_1\rangle_\rho|
+|\langle V_2\rangle_\rho|
+|\langle V_3\rangle_\rho|
\Bigr).
\end{aligned}
\label{eq:main_additive}
\end{equation}
In the following, we will present a proof of the above inequality. We assume throughout that all variances and commutator expectation values appearing in the derivation are finite and well defined. The proof therefore applies directly to finite-dimensional systems and bounded Hermitian observables.

\paragraph*{Proof.}
Introduce the vector operators
\begin{equation}
\mathbf{L}=(L_1,L_2,L_3),\qquad
\mathbf{V}=(V_1,V_2,V_3),
\end{equation}
so that Eq.~\eqref{eq:commutator_form} can be written compactly as
\begin{equation}
\mathbf{L}\times\mathbf{L}=\mathrm{i}\mathbf{V},
\end{equation}
here the cross product is understood componentwise, so that, for example,
$(\mathbf L\times \mathbf L)_1=[L_2,L_3]$. If
\begin{equation}
|\langle V_1\rangle_\rho|+|\langle V_2\rangle_\rho|+|\langle V_3\rangle_\rho|=0,
\end{equation}
then the right-hand side of Eq.~\eqref{eq:main_additive} vanishes, and the inequality
is trivial. Hence it suffices to consider the case $\langle\mathbf{V}\rangle_\rho\neq 0$.

We first note that
\begin{equation}
\begin{split}
(\Delta_\rho L_1)^2+(\Delta_\rho L_2)^2+(\Delta_\rho L_3)^2
&=
\langle \mathbf{L}^2\rangle_\rho
-|\langle \mathbf{L}\rangle_\rho|^2,
\end{split}
\label{eq:rot_inv_num}
\end{equation}
which is invariant under any proper rotation acting simultaneously on the
components of $\mathbf{L}$.
Moreover,
\begin{equation}
|\langle V_1\rangle_\rho|+|\langle V_2\rangle_\rho|+|\langle V_3\rangle_\rho|
\le
\sqrt{3}\,|\langle \mathbf{V}\rangle_\rho|,
\label{eq:cauchy_step}
\end{equation}
where equality holds if and only if
\begin{equation}
|\langle V_1\rangle_\rho|
=
|\langle V_2\rangle_\rho|
=
|\langle V_3\rangle_\rho|.
\label{eq:equalV}
\end{equation}
Therefore,
\begin{equation}
\begin{aligned}
\frac{
(\Delta_\rho L_1)^2+(\Delta_\rho L_2)^2+(\Delta_\rho L_3)^2
}{
|\langle V_1\rangle_\rho|+|\langle V_2\rangle_\rho|
+|\langle V_3\rangle_\rho|
}
\ge \\
\frac{
(\Delta_\rho L_1)^2+(\Delta_\rho L_2)^2+(\Delta_\rho L_3)^2
}{
\sqrt{3}\,|\langle \mathbf{V}\rangle_\rho|
}.
\end{aligned}
\label{eq:first_ratio_step}
\end{equation}
Now choose a rotated coordinate system $(x',y',z')$ such that
$\langle\mathbf{V}\rangle_\rho$ is parallel to the $z'$ axis.
Denote the rotated components of $\mathbf{L}$ and $\mathbf{V}$ by
$L'_x,L'_y,L'_z$ and $V'_x,V'_y,V'_z$, respectively.
Then, by Eq.~\eqref{eq:rot_inv_num},
\begin{equation}
\begin{aligned}
(\Delta_\rho L_1)^2+(\Delta_\rho L_2)^2+(\Delta_\rho L_3)^2
=\\
(\Delta_\rho L'_x)^2+(\Delta_\rho L'_y)^2+(\Delta_\rho L'_z)^2,
\end{aligned}
\end{equation}
and
\begin{equation}
|\langle \mathbf{V}\rangle_\rho|=|\langle V'_z\rangle_\rho|.
\end{equation}
Hence,
\begin{align}
&\frac{
(\Delta_\rho L_1)^2+(\Delta_\rho L_2)^2+(\Delta_\rho L_3)^2
}{
|\langle V_1\rangle_\rho|+|\langle V_2\rangle_\rho|+|\langle V_3\rangle_\rho|
}
\notag\\
&\ge
\frac{
(\Delta_\rho L'_x)^2+(\Delta_\rho L'_y)^2+(\Delta_\rho L'_z)^2
}{
\sqrt{3}\,|\langle V'_z\rangle_\rho|
}
\notag\\
&\ge
\frac{
2\Delta_\rho L'_x\,\Delta_\rho L'_y+(\Delta_\rho L'_z)^2
}{
\sqrt{3}\,|\langle V'_z\rangle_\rho|
}.
\label{eq:agm_step}
\end{align}
Next, by the Robertson uncertainty relation applied to $L'_x$ and $L'_y$,
\begin{equation}
\Delta_\rho L'_x\,\Delta_\rho L'_y
\ge
\frac{1}{2}\bigl|\langle [L'_x,L'_y]\rangle_\rho\bigr|
=
\frac{1}{2}|\langle V'_z\rangle_\rho|.
\label{eq:robertson_xy}
\end{equation}
Substituting Eq.~\eqref{eq:robertson_xy} into Eq.~\eqref{eq:agm_step}, we obtain
\begin{equation}
\begin{split}
\frac{
(\Delta_\rho L_1)^2+(\Delta_\rho L_2)^2+(\Delta_\rho L_3)^2
}{
|\langle V_1\rangle_\rho|+|\langle V_2\rangle_\rho|+|\langle V_3\rangle_\rho|
}
&\ge
\frac{
|\langle V'_z\rangle_\rho|+(\Delta_\rho L'_z)^2
}{
\sqrt{3}\,|\langle V'_z\rangle_\rho|
}
\\[1pt]
&\ge
\frac{1}{\sqrt{3}}.
\end{split}
\label{eq:last_step}
\end{equation}
This proves Eq.~\eqref{eq:main_additive}.
\hfill $\square$

\medskip

We now characterize the equality condition in Eq.~\eqref{eq:main_additive}
for a general density operator $\rho$.
Let $\Pi_\rho$ denote the support projector of $\rho$, namely, the orthogonal
projection onto $\operatorname{supp}(\rho)$. We summarize the conclusion as the following theorem.

\begin{theorem}[Equality condition]
\label{thm:equality-condition}

Equality in Eq.~\eqref{eq:main_additive} holds if and only if one of the following
two cases occurs.
\end{theorem}
\paragraph*{(i) Trivial branch.}
The right-hand side of Eq.~\eqref{eq:main_additive} vanishes, namely
\begin{equation}
|\langle V_1\rangle_\rho|+|\langle V_2\rangle_\rho|+|\langle V_3\rangle_\rho|=0,
\label{eq:trivial_rhs_zero}
\end{equation}
and simultaneously
\begin{equation}
(\Delta_\rho L_1)^2=(\Delta_\rho L_2)^2=(\Delta_\rho L_3)^2=0.
\label{eq:trivial_var_zero}
\end{equation}
Equivalently,
\begin{equation}
\begin{split}
(L_j-\langle L_j\rangle_\rho I)\Pi_\rho=0,
\qquad j=1,2,3.
\end{split}
\label{eq:trivial_support}
\end{equation}
\paragraph*{(ii) Nontrivial branch.}
Suppose that
\begin{equation}
|\langle V_1\rangle_\rho|+|\langle V_2\rangle_\rho|+|\langle V_3\rangle_\rho|\neq 0.
\end{equation}
Then equality in Eq.~\eqref{eq:main_additive} holds if and only if all of the
following conditions are satisfied:

\begin{enumerate}
\item
Equality must hold in Eq.~\eqref{eq:cauchy_step}. Hence
\begin{equation}
|\langle V_1\rangle_\rho|
=
|\langle V_2\rangle_\rho|
=
|\langle V_3\rangle_\rho|.
\label{eq:gen_equalV}
\end{equation}

\item
Equality must hold in the second step of  Eq.~\eqref{eq:agm_step}. Hence
\begin{equation}
\Delta_\rho L'_x=\Delta_\rho L'_y.
\label{eq:gen_equal_var}
\end{equation}

\item
Equality must hold in Eq.~\eqref{eq:robertson_xy}. Therefore, by the equality
condition of the Robertson uncertainty relation for general density operators
(see Appendix~\ref{app:robertson_equality}), there exists a nonzero real number
$\lambda$ such that
\begin{equation}
\Bigl[
(L'_x-\langle L'_x\rangle_\rho I)
+i\lambda (L'_y-\langle L'_y\rangle_\rho I)
\Bigr]\Pi_\rho=0.
\label{eq:gen_robertson_support}
\end{equation}

\item
Equality must hold in the last inequality of Eq.~\eqref{eq:last_step}. Hence
\begin{equation}
(\Delta_\rho L'_z)^2=0,
\label{eq:gen_Lz_var_zero}
\end{equation}
or equivalently,
\begin{equation}
(L'_z-\langle L'_z\rangle_\rho I)\Pi_\rho=0.
\label{eq:gen_Lz_support}
\end{equation}
\end{enumerate}

Conversely, if either the trivial branch or the nontrivial branch holds, then
every intermediate inequality in the above proof is saturated, and therefore
equality in Eq.~\eqref{eq:main_additive} follows.
This gives a necessary and sufficient condition for the saturation of
Eq.~\eqref{eq:main_additive}.

\section{Geometric meaning and interpretation of the equality condition}
\label{sec:geometric_meaning}

The above derivation shows that Eq.~\eqref{eq:main_additive} arises from two complementary ingredients: a rotationally invariant uncertainty constraint and a geometric inequality involving the norm of the commutator vector. To characterize the saturation condition, we now introduce the covariance matrix $\Gamma(\rho)$ of the observable triple $(L_1,L_2,L_3)$, which is defined and discussed in Appendix~\ref{app:covariance_geometry}. Then
\begin{equation}
\operatorname{tr}\Gamma(\rho)
=
(\Delta_\rho L_1)^2+(\Delta_\rho L_2)^2+(\Delta_\rho L_3)^2 .
\label{eq:trace_cov_total_variance}
\end{equation}
The argument in the proof gives the sharper rotationally invariant estimate
\begin{equation}
\operatorname{tr}\Gamma(\rho)
\geq
|\langle\mathbf V\rangle_\rho|.
\label{eq:rot_inv_uncertainty_bound}
\end{equation}
Together with the elementary norm comparison in $\mathbb R^3$,
\begin{equation}
|\langle\mathbf V\rangle_\rho|
\geq
\frac{1}{\sqrt{3}}
\|\langle\mathbf V\rangle_\rho\|_1,
\label{eq:l2_l1_comparison}
\end{equation}
where
\begin{equation}
\|\langle\mathbf V\rangle_\rho\|_1
=
|\langle V_1\rangle_\rho|
+
|\langle V_2\rangle_\rho|
+
|\langle V_3\rangle_\rho|,
\end{equation}
one obtains
\begin{equation}
\operatorname{tr}\Gamma(\rho)
\geq
|\langle\mathbf V\rangle_\rho|
\geq
\frac{1}{\sqrt{3}}
\|\langle\mathbf V\rangle_\rho\|_1 .
\label{eq:core_geometric_chain}
\end{equation}
Thus the constant $1/\sqrt{3}$  comes from the
comparison between the Euclidean norm and the $\ell^1$ norm of the commutator
vector $\langle\mathbf V\rangle_\rho$. The genuinely quantum part of the
inequality is the rotationally invariant bound
\eqref{eq:rot_inv_uncertainty_bound}, while the passage from
\eqref{eq:rot_inv_uncertainty_bound} to Eq.~\eqref{eq:main_additive} is a
coordinate-dependent geometric step.

\begin{figure}[t]
\centering
\includegraphics[width=1.0\columnwidth]{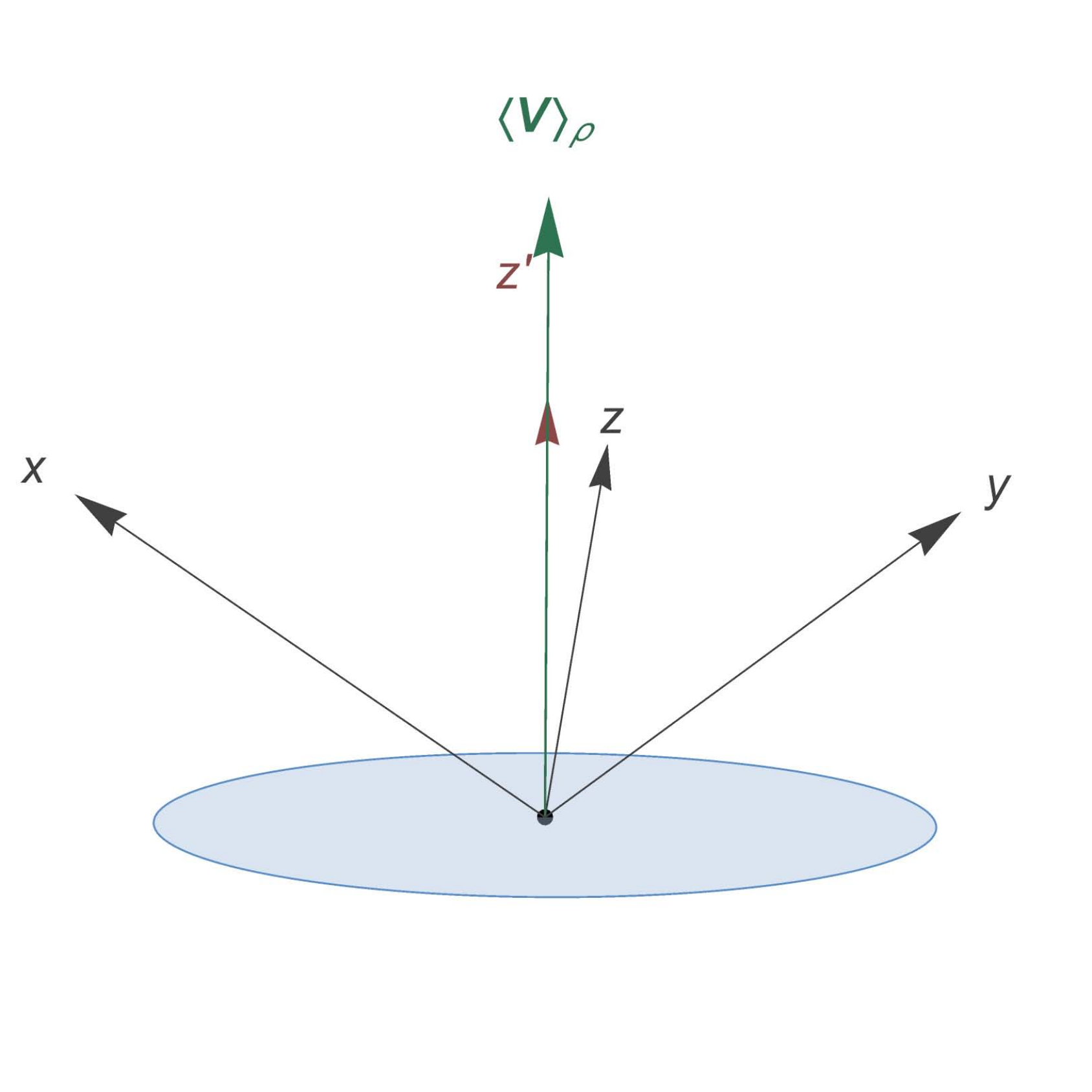}
\caption{
Geometric illustration of the equality condition in the additive
triple-observable uncertainty relation. The vector
$\langle\mathbf V\rangle_\rho
=(\langle V_1\rangle_\rho,\langle V_2\rangle_\rho,\langle V_3\rangle_\rho)$
points along one of the body diagonals of the original $(x,y,z)$ frame, namely
$|\langle V_1\rangle_\rho|
=|\langle V_2\rangle_\rho|
=|\langle V_3\rangle_\rho|$.
In the adapted coordinate system with
$z'\parallel \langle\mathbf V\rangle_\rho$, the covariance ellipsoid
degenerates into a disk in the transverse $x'y'$-plane.
}
\label{fig:covariance_disk}
\end{figure}

The equality condition admits a simple geometric interpretation. In the
nontrivial equality case, Eq.~\eqref{eq:gen_equalV} says that
$\langle\mathbf V\rangle_\rho$ points along one of the body diagonals of the
original coordinate frame. Equivalently, in the adapted coordinate system
$(x',y',z')$ used in the proof, the $z'$-axis is chosen parallel to
$\langle\mathbf V\rangle_\rho$.
Moreover, Eq.~\eqref{eq:gen_Lz_var_zero} shows that the variance of the
projected observable in the direction of $\langle\mathbf V\rangle_\rho$
vanishes. Hence there is no longitudinal fluctuation in the equality case.
The remaining uncertainty is confined to the transverse plane orthogonal to
$\langle\mathbf V\rangle_\rho$. By Eq.~\eqref{eq:gen_equal_var} and the
saturation of Eq.~\eqref{eq:robertson_xy}, the two transverse variances are
equal and attain the Robertson lower bound. Therefore, as illustrated in
Fig.~\ref{fig:covariance_disk}, the covariance ellipsoid degenerates into a
disk lying in the plane perpendicular to $\langle\mathbf V\rangle_\rho$.

In this sense, the equality case separates the total uncertainty into a
longitudinal part and a transverse part. The longitudinal part is absent in the
saturating configuration, while the transverse part is fixed by the
noncommutativity encoded in $\langle\mathbf V\rangle_\rho$. The equality case
therefore represents the situation in which all unavoidable uncertainty is
concentrated in the plane orthogonal to the commutator vector.

\section{Application: an inverse construction from prescribed states}
\label{sec:inverse_construction}

Instead of fixing the observables first and then determining which states
saturate the inequality, one may ask the converse question: given prescribed
states, can one construct a triple of observables for which the nontrivial
equality in Eq.~\eqref{eq:main_additive} is attained? There are some interesting
consequences of this inverse problem associated with Eq.~\eqref{eq:main_additive}.

The equality conditions obtained above suggest that it is natural to work first
in an adapted coordinate system $(x',y',z')$. In such a frame, the inverse
problem amounts to constructing a triple $(L'_x,L'_y,L'_z)$ such that the
prescribed states satisfy the adapted-frame part of the nontrivial equality
conditions, namely Eqs.~\eqref{eq:gen_equal_var},
\eqref{eq:gen_robertson_support}, and \eqref{eq:gen_Lz_support}. Once such a
triple has been found and $\langle\mathbf V'\rangle_\rho$ is parallel to the
$z'$-axis, one may pass back to the original coordinates by choosing an
orthogonal matrix $O\in SO(3)$ such that
\begin{equation}
O
\begin{pmatrix}
0\\
0\\
1
\end{pmatrix}
=
\frac{1}{\sqrt{3}}
\begin{pmatrix}
1\\
1\\
1
\end{pmatrix},
\label{eq:inverse_rotation_condition}
\end{equation}
and defining
\begin{equation}
\begin{pmatrix}
L_1\\
L_2\\
L_3
\end{pmatrix}
=
O
\begin{pmatrix}
L'_x\\
L'_y\\
L'_z
\end{pmatrix}.
\label{eq:inverse_rotation_back}
\end{equation}
After this rotation, the vector $\langle\mathbf V\rangle_\rho$ points along a
body-diagonal direction of the original coordinate frame, so that
Eq.~\eqref{eq:gen_equalV} is also satisfied. Hence the state saturates
Eq.~\eqref{eq:main_additive} for the triple $(L_1,L_2,L_3)$.

A simple realization of this idea is provided by finite-dimensional
representations of $\mathfrak{su}(2)$. Let $\mathcal K$ be an $\ell$-dimensional Hilbert
space, with $\ell=2j+1$, and choose an orthonormal basis
\begin{equation}
\{u_m\}_{m=-j}^{j}.
\label{eq:weight_basis}
\end{equation}
On this basis, take the standard spin-$j$ irreducible representation with
generators $J_x,J_y,J_z$ satisfying
\begin{equation}
[J_x,J_y]=\mathrm{i}J_z,\qquad
[J_y,J_z]=\mathrm{i}J_x,\qquad
[J_z,J_x]=\mathrm{i}J_y,
\label{eq:su2_commutation}
\end{equation}
and
\begin{equation}
J_z u_m=m u_m,\qquad m=-j,-j+1,\ldots,j.
\label{eq:Jz_weight_basis}
\end{equation}
Set
\begin{equation}
L'_x=J_x,\qquad
L'_y=J_y,\qquad
L'_z=J_z.
\label{eq:su2_primed_generators}
\end{equation}

We now show that, in this irreducible representation, the only states
satisfying the adapted-frame part of the nontrivial equality conditions are the
two extremal weight pure states $u_j$ and $u_{-j}$. Indeed,
Eq.~\eqref{eq:gen_Lz_support} requires
\begin{equation}
(J_z-\langle J_z\rangle_\rho I)\Pi_\rho=0.
\label{eq:Jz_support_condition_inverse}
\end{equation}
Since the spectrum of $J_z$ in the irreducible spin-$j$ representation is
simple, a density operator satisfying this condition must be supported on one
of the one-dimensional eigenspaces of $J_z$. Hence it must be the rank-one
projector onto one of the weight vectors $u_m$. For the pure state $u_m$, one has
\begin{equation}
\langle J_z\rangle_{u_m}=m,\qquad
\Delta^2_{u_m}J_z=0,
\label{eq:Jz_weight_variance_inverse}
\end{equation}
and
\begin{equation}
\Delta^2_{u_m}J_x
=
\Delta^2_{u_m}J_y
=
\frac{1}{2}\bigl(j(j+1)-m^2\bigr).
\label{eq:Jx_Jy_weight_variance_inverse}
\end{equation}
Thus Eq.~\eqref{eq:gen_equal_var} is automatically satisfied for every weight
pure state. The remaining Robertson saturation condition
\eqref{eq:gen_robertson_support}, or equivalently saturation of
Eq.~\eqref{eq:robertson_xy}, gives
\begin{equation}
\frac{1}{2}\bigl(j(j+1)-m^2\bigr)
=
\frac{1}{2}|m|.
\label{eq:weight_saturation_condition_inverse}
\end{equation}
Equivalently,
\begin{equation}
j(j+1)-m^2=|m|.
\label{eq:weight_saturation_equation_inverse}
\end{equation}
If $m\geq 0$, this becomes
\begin{equation}
(j-m)(j+m+1)=0,
\end{equation}
and hence $m=j$. If $m\leq 0$, it becomes
\begin{equation}
(j+m)(j-m+1)=0,
\end{equation}
and hence $m=-j$. Therefore the only  states satisfying the adapted-frame
part of the nontrivial equality conditions in the irreducible block are pure states
$u_j$ and $u_{-j}$.

Consequently, for any two orthogonal pure states $|\phi_1\rangle$ and
$|\phi_2\rangle$, one may choose an orthonormal basis of $\mathcal K$ such that
\begin{equation}
u_j=|\phi_1\rangle,\qquad
u_{-j}=|\phi_2\rangle.
\label{eq:assign_prescribed_states_inverse}
\end{equation}
The construction above then gives a triple $(L'_x,L'_y,L'_z)$ for which the two
prescribed pure states are precisely the pure states satisfying the adapted-frame
part of the nontrivial equality conditions. Applying the rotation in
Eqs.~\eqref{eq:inverse_rotation_condition} and
\eqref{eq:inverse_rotation_back}, one obtains a triple of Hermitian operators
$(L_1,L_2,L_3)$ for which the prescribed pure states $|\phi_1\rangle$ and
$|\phi_2\rangle$ saturate Eq.~\eqref{eq:main_additive}.

More generally, reducible representations provide further solutions of the
inverse problem. If several irreducible blocks have the same highest weight
or the same lowest weight, then the corresponding extremal eigenspace of
$J_z$ is degenerate. In this case, every pure state in that extremal eigenspace
satisfies the adapted-frame equality conditions. Moreover, every density
operator supported on this extremal eigenspace also satisfies them, since the
support condition \eqref{eq:gen_Lz_support} and the Robertson support condition
\eqref{eq:gen_robertson_support} hold on the whole support.

As a simple illustration of the inverse construction introduced above, consider a two-qubit
system and choose the ordered Bell basis
\begin{equation}
\left\{
\frac{|00\rangle-|11\rangle}{\sqrt{2}},\,
\frac{|01\rangle-|10\rangle}{\sqrt{2}},\,
\frac{|01\rangle+|10\rangle}{\sqrt{2}},\,
\frac{|00\rangle+|11\rangle}{\sqrt{2}}
\right\}.
\label{eq:bell_basis}
\end{equation}
With respect to this basis, one convenient spin-$3/2$ triple is given by
\begin{equation}
L'_x=
\frac{\sqrt{3}}{2}\sigma_{IX}
+\frac{1}{2}\sigma_{IZ}
-\frac{1}{2}\sigma_{ZI},
\label{eq:two_qubit_Lxprime}
\end{equation}
\begin{equation}
L'_y=
-\frac{1}{2}\sigma_{XY}
+\frac{1}{2}\sigma_{YX}
+\frac{\sqrt{3}}{2}\sigma_{YZ},
\label{eq:two_qubit_Lyprime}
\end{equation}
and
\begin{equation}
L'_z=
\sigma_{XX}
-\frac{1}{2}\sigma_{YY},
\label{eq:two_qubit_Lzprime}
\end{equation}
where $\sigma_{AB}:=\sigma_A\otimes\sigma_B$ with
$A,B\in\{I,X,Y,Z\}$. These operators satisfy the $su(2)$ commutation relations.
In the Bell basis of Eq.~\eqref{eq:bell_basis}, one has
\begin{equation}
L'_z=
\operatorname{diag}\left(
-\frac{3}{2},-\frac{1}{2},\frac{1}{2},\frac{3}{2}
\right).
\label{eq:Lzprime_bell_diagonal}
\end{equation}
Hence the first and the last Bell states are respectively the lowest-weight
and highest-weight states. By the argument above, these two Bell states satisfy
the adapted-frame equality conditions for the primed triple. The corresponding
observables in the original coordinate system are then obtained by the rotation
procedure described above.

The construction is also stable under local unitary transformations. Indeed,
if \(U_{\mathrm{loc}}\) is a local unitary operator and
\[
\widetilde{L}'_\alpha
=
U_{\mathrm{loc}} L'_\alpha U_{\mathrm{loc}}^\dagger,
\qquad \alpha=x,y,z,
\]
then state $U_{\mathrm{loc}} \rho U_{\mathrm{loc}}^\dagger$ satisfies the adapted-frame equality
conditions for the transformed triple
\((\widetilde{L}'_x,\widetilde{L}'_y,\widetilde{L}'_z)\) whenever
$\rho $ satisfies them for \((L'_x,L'_y,L'_z)\). In particular, in
the two-qubit case, since every maximally entangled pure state is locally
unitarily equivalent to a Bell state, an appropriate choice of
\(U_{\mathrm{loc}}\) gives a triple of observables for which any prescribed
maximally entangled two-qubit state saturates the inequality.

The systematic construction described above is not restricted to the
two-qubit example. It can in principle be applied to different
finite-dimensional systems, including, for instance, a pair of orthogonal
GHZ states in an \(n\)-qubit Hilbert space. The explicit form of the
resulting observable triple, however, may become considerably more
complicated as the dimension increases.

\section{Discussion and Conclusion}

We have studied an additive uncertainty relation for three Hermitian observables from the viewpoint of rotational symmetry and covariance geometry. The rotational proof reduces the inequality to the Robertson uncertainty relation for a transverse pair, together with a geometric comparison of norms. This gives a necessary and sufficient condition for saturation for general density operators. Geometrically, the nontrivial equality case is characterized by a disk-like degeneration of the covariance ellipsoid in the plane perpendicular to the commutator vector. Thus the equality condition shows how the unavoidable uncertainty is distributed between the longitudinal and transverse directions. We have also presented an inverse construction based on finite-dimensional representations of $\mathfrak{su}(2)$, which provides triples of observables with prescribed saturating states.
These results clarify the structure of tight additive uncertainty relations and provide a systematic way to construct and interpret equality cases in finite-dimensional quantum systems.

\begin{acknowledgments}
J.L.C. is supported by the Quantum Science and Technology-National Science and Technology Major Project (Grant No. 2024ZD0301000), and the National Natural Science Foundation of China (Grant No. 12275136).
      Y.Y.Z. is supported by the Nankai Zhide Foundation.
	
\end{acknowledgments}

\appendix
\section{Robertson uncertainty relation and its equality condition}
\label{app:robertson_equality}

In this appendix we recall the Robertson uncertainty relation and the equality
condition used in Sec.~\ref{sec:additive_relation}. Let $A$ and $B$ be two
Hermitian operators and let $\rho$ be a density operator. Define the centered
observables
\begin{equation}
\widetilde A:=A-\langle A\rangle_\rho I,
\qquad
\widetilde B:=B-\langle B\rangle_\rho I.
\end{equation}
Then
\begin{equation}
(\Delta_\rho A)^2=\langle \widetilde A^2\rangle_\rho,
\qquad
(\Delta_\rho B)^2=\langle \widetilde B^2\rangle_\rho.
\end{equation}

For every real number $t$, consider
\begin{equation}
Q(t)
:=
\operatorname{tr}\rho
(\widetilde A+i t\widetilde B)^\dagger
(\widetilde A+i t\widetilde B).
\end{equation}
Since $Q(t)$ is a squared Hilbert--Schmidt norm,
\begin{equation}
Q(t)
=
\|(\widetilde A+i t\widetilde B)\sqrt{\rho}\|_{\mathrm{HS}}^2
\geq 0.
\end{equation}
A direct expansion gives
\begin{equation}
Q(t)
=
(\Delta_\rho A)^2
+t^2(\Delta_\rho B)^2
+i t\langle [A,B]\rangle_\rho.
\end{equation}
Because $[A,B]$ is anti-Hermitian, the number
\begin{equation}
\gamma:=-i\langle [A,B]\rangle_\rho
\end{equation}
is real. Hence
\begin{equation}
Q(t)
=
(\Delta_\rho A)^2
+t^2(\Delta_\rho B)^2
-t\gamma.
\end{equation}
Since this quadratic polynomial is nonnegative for all real $t$, its
discriminant is nonpositive:
\begin{equation}
\gamma^2
\leq
4(\Delta_\rho A)^2(\Delta_\rho B)^2.
\end{equation}
Equivalently,
\begin{equation}
\Delta_\rho A\,\Delta_\rho B
\geq
\frac{1}{2}
|\langle[A,B]\rangle_\rho|.
\label{eq:appendix_robertson}
\end{equation}
This is the Robertson uncertainty relation.

We now record the equality condition in the nontrivial case
\begin{equation}
|\langle[A,B]\rangle_\rho|\neq 0.
\end{equation}
Then both $\Delta_\rho A$ and $\Delta_\rho B$ are nonzero. Equality in
Eq.~\eqref{eq:appendix_robertson} holds if and only if the quadratic polynomial
$Q(t)$ has a real zero. This occurs if and only if
\begin{equation}
Q(\lambda)=0
\end{equation}
for
\begin{equation}
\lambda
=
\frac{\gamma}{2(\Delta_\rho B)^2}
=
\frac{-i\langle[A,B]\rangle_\rho}
{2(\Delta_\rho B)^2}.
\label{eq:lambda_robertson}
\end{equation}
Since
\begin{equation}
Q(\lambda)
=
\|(\widetilde A+i\lambda\widetilde B)\sqrt{\rho}\|_{\mathrm{HS}}^2,
\end{equation}
the condition $Q(\lambda)=0$ is equivalent to
\begin{equation}
(\widetilde A+i\lambda\widetilde B)\sqrt{\rho}=0.
\end{equation}
Let $\Pi_\rho$ be the support projector of $\rho$. Since the range of
$\sqrt{\rho}$ is $\operatorname{supp}(\rho)$, this is equivalent to
\begin{equation}
(\widetilde A+i\lambda\widetilde B)\Pi_\rho=0.
\label{eq:appendix_robertson_support}
\end{equation}
Thus, in the nontrivial case, equality in the Robertson uncertainty relation
holds if and only if there exists a nonzero real number $\lambda$ such that
\begin{equation}
\Bigl[
A-\langle A\rangle_\rho I
+i\lambda(B-\langle B\rangle_\rho I)
\Bigr]\Pi_\rho=0.
\label{eq:appendix_robertson_equality_condition}
\end{equation}

Conversely, if Eq.~\eqref{eq:appendix_robertson_equality_condition} holds for a
real nonzero $\lambda$, then
\begin{equation}
\widetilde A\sqrt{\rho}
=
-i\lambda\,\widetilde B\sqrt{\rho}.
\end{equation}
Therefore
\begin{equation}
(\Delta_\rho A)^2
=
\lambda^2(\Delta_\rho B)^2,
\label{eq:variance_ratio_lambda}
\end{equation}
and
\begin{equation}
\langle \widetilde A\widetilde B\rangle_\rho
=
i\lambda(\Delta_\rho B)^2.
\end{equation}
In particular, the symmetric covariance vanishes:
\begin{equation}
\frac{1}{2}
\langle \widetilde A\widetilde B+\widetilde B\widetilde A\rangle_\rho
=0,
\end{equation}
and
\begin{equation}
\langle[A,B]\rangle_\rho
=
2i\lambda(\Delta_\rho B)^2.
\end{equation}
It follows that
\begin{equation}
\frac{1}{2}|\langle[A,B]\rangle_\rho|
=
|\lambda|(\Delta_\rho B)^2
=
\Delta_\rho A\,\Delta_\rho B.
\end{equation}
Hence the Robertson inequality is saturated.

Finally, Eq.~\eqref{eq:variance_ratio_lambda} shows that if, in addition,
\begin{equation}
\Delta_\rho A=\Delta_\rho B\neq 0,
\end{equation}
then
\begin{equation}
|\lambda|=1.
\end{equation}
This is the implication used in Eq.~\eqref{eq:gen_robertson_support} after
Eq.~\eqref{eq:gen_equal_var}.

\section{Covariance matrix and covariance ellipsoid}
\label{app:covariance_geometry}

For the triple $(L_1,L_2,L_3)$ and the state $\rho$, define the centered
observables
\begin{equation}
\widetilde L_i:=L_i-\langle L_i\rangle_\rho I,
\qquad i=1,2,3.
\end{equation}
The covariance matrix is the real symmetric matrix
\begin{equation}
\Gamma(\rho)_{ij}
:=
\frac{1}{2}
\Bigl\langle
\widetilde L_i\widetilde L_j+\widetilde L_j\widetilde L_i
\Bigr\rangle_\rho,
\qquad i,j=1,2,3.
\label{eq:covariance_matrix}
\end{equation}
It is positive semidefinite, and for every unit vector
$\mathbf n=(n_1,n_2,n_3)^T\in\mathbb R^3$, the variance of the projected observable
$\mathbf n\cdot\mathbf L$ is
\begin{equation}
\Delta_\rho^2(\mathbf n\cdot\mathbf L)
=
n^T\Gamma(\rho)n.
\label{eq:directional_variance}
\end{equation}
Thus $\Gamma(\rho)$ describes the directional variances of the triple
$(L_1,L_2,L_3)$.

The associated covariance ellipsoid is defined by
\begin{equation}
\mathcal E(\rho):=
\Gamma(\rho)^{1/2}(B),
\label{eq:covariance_ellipsoid}
\end{equation}
where
\begin{equation}
B=\{u\in\mathbb R^3:\ |u|\leq 1\}
\end{equation}
is the unit ball. This ellipsoid gives a geometric representation of the
uncertainty of the triple in different directions.

In the adapted coordinate system used in the proof of
Eq.~\eqref{eq:main_additive}, the nontrivial equality condition implies
\begin{equation}
\Delta_\rho^2 L'_z=0,
\end{equation}
and hence all covariances involving $L'_z$ vanish. Furthermore,
Eq.~\eqref{eq:gen_equal_var} gives
\begin{equation}
\Delta_\rho^2 L'_x=\Delta_\rho^2 L'_y,
\end{equation}
while the Robertson equality condition
\eqref{eq:gen_robertson_support} implies that the symmetric covariance between
$L'_x$ and $L'_y$ vanishes. Therefore, in the equality case the covariance
matrix in the adapted frame takes the form
\begin{equation}
\Gamma'(\rho)
=
\begin{pmatrix}
r^2 & 0 & 0\\
0 & r^2 & 0\\
0 & 0 & 0
\end{pmatrix},
\label{eq:equality_covariance_disk_matrix}
\end{equation}
with
\begin{equation}
r^2=\frac{1}{2}|\langle V'_z\rangle_\rho|.
\end{equation}
Consequently, the covariance ellipsoid degenerates into a disk in the
$x'y'$-plane.

\end{document}